\documentclass[twocolumn]{aastex631}

\shorttitle{A Perfect Tidal Storm}
\shortauthors{Stephen R. Kane et al.}

\begin{document}

\title{A Perfect Tidal Storm: HD 104067 Planetary Architecture
  Creating an Incandescent World}

\author[0000-0002-7084-0529]{Stephen R. Kane}
\affiliation{Department of Earth and Planetary Sciences, University of
  California, Riverside, CA 92521, USA}
\email{skane@ucr.edu}

\author[0000-0002-3551-279X]{Tara Fetherolf}
\affiliation{Department of Physics, California State University, San
  Marcos, CA 92096, USA}
\affiliation{Department of Earth and Planetary Sciences, University of
  California, Riverside, CA 92521, USA}

\author[0000-0002-4860-7667]{Zhexing Li}
\affiliation{Department of Earth and Planetary Sciences, University of
  California, Riverside, CA 92521, USA}

\author[0000-0001-7047-8681]{Alex S. Polanski}
\affiliation{Department of Physics and Astronomy, University of
  Kansas, Lawrence, KS 66045, USA}

\author[0000-0001-8638-0320]{Andrew W. Howard}
\affiliation{Department of Astronomy, California Institute of
  Technology, Pasadena, CA 91125, USA}

\author[0000-0002-0531-1073]{Howard Isaacson}
\affiliation{Department of Astronomy, University of California,
  Berkeley, CA 94720, USA}
\affiliation{Centre for Astrophysics, University of Southern
  Queensland, Toowoomba, QLD 4350, Australia}

\author[0000-0003-4603-556X]{Teo Mo\v{c}nik}
\affiliation{Gemini Observatory/NSF's NOIRLab, 670 N. A'ohoku Place,
  Hilo, HI 96720, USA}

\author[0009-0000-4164-2358]{Sadie G. Welter}
\affiliation{Department of Earth and Planetary Sciences, University of
  California, Riverside, CA 92521, USA}

%%%%%%%%%%%%%%%%%%%%%%%%%%%%%%%%%%%%%%%%%%%%%%%%%%%%%%%%%%%%%%%%%%%%

\begin{abstract}

The discovery of planetary systems beyond the solar system has
revealed a diversity of architectures, most of which differ
significantly from our system. The initial detection of an exoplanet
is often followed by subsequent discoveries within the same system as
observations continue, measurement precision is improved, or
additional techniques are employed. The HD~104067 system is known to
consist of a bright K dwarf host star and a giant planet in a
$\sim$55~day period eccentric orbit. Here we report the discovery of
an additional planet within the HD~104067 system, detected through the
combined analysis of radial velocity data from the HIRES and HARPS
instruments. The new planet has a mass similar to Uranus and is in an
eccentric $\sim$14 day orbit. Our injection-recovery analysis of the
radial velocity data exclude Saturn-mass and Jupiter-mass planets out
to 3~AU and 8~AU, respectively. We further present TESS observations
that reveal a terrestrial planet candidate ($R_p =
1.30\pm0.12$~$R_\oplus$) in a $\sim$2.2~day period orbit. Our
dynamical analysis of the three planet model shows that the two outer
planets produce significant eccentricity excitation of the inner
planet, resulting in tidally induced surface temperatures as high as
$\sim$2600~K for an emissivity of unity. The terrestrial planet
candidate may therefore be caught in a tidal storm, potentially
resulting in its surface radiating at optical wavelengths.

\end{abstract}


%%%%%%%%%%%%%%%%%%%%%%%%%%%%%%%%%%%%%%%%%%%%%%%%%%%%%%%%%%%%%%%%%%%%

\section{Introduction}
\label{intro}

In the space of only a few decades, the exoplanet inventory has
dramatically grown from zero to over 5000. The vast majority of these
contributions to the exoplanet inventory have originated from the
radial velocity (RV) and transit methods of exoplanet detection. RV
surveys are increasing in both measurement precision
\citep{fischer2016} and survey duration, extending their sensitivity
to well beyond the snow line
\citep{wittenmyer2020b,fulton2021,bonomo2023} and enabling numerous
comparisons to the solar system architecture
\citep{martin2015b,horner2020b,raymond2020a,kane2021d}. Discoveries
via the transit method have primarily arrived via the Kepler mission
\citep{borucki2016} and the Transiting Exoplanet Survey Satellite
(TESS; \citet{ricker2015}). The vast number of exoplanet discoveries
have opened up new explorations of planetary system architectures
\citep{ford2014,winn2015,he2019,mishra2023a} and the dynamical
evolution that has shaped these configurations
\citep{rasio1996c,juric2008b,ida2013,kane2014b}. TESS observations
have included those of previously known planetary systems
\citep{kane2021b}, allowing the realization that some of these
RV-detected planets do in fact transit \citep{kane2020c,pepper2020}
and also have additional planetary companions
\citep{huang2018,teske2020}. Continued observations of the known
systems also improve the orbits of the planets contained therein,
updating the orbital ephemerides and planetary bulk properties
\citep{kane2009c,dragomir2020a}. Each new discovery within these known
systems adds to their complexity, and creates further opportunities to
provide dynamical constraints that may inform follow-up observations.

One such known planetary system, HD~104067, consists of a bright ($V
\sim 8$) early K dwarf host star located $\sim$20~pcs
away. \citet{segransan2011} announced the discovery of a planet
orbiting HD~104067 using RV data obtained with the High Accuracy
Radial velocity Planet Searcher (HARPS) spectrograph
\citep{pepe2000}. The planet was found to have a mass of
$\sim$0.2~$M_J$ and an orbital period of $\sim$55~days, and was
assumed to have a circular orbit. \citet{rosenthal2021} independently
confirmed the presence of the planet using the High Resolution Echelle
Spectrometer (HIRES) on the Keck I telescope \citep{vogt1994},
although their Keplerian model preferred an eccentric orbit ($e =
0.25$) for the planet. In the meantime, RV observations of the system
continued and TESS has observed the star during several sectors of
observation, obtaining high precision photometry for the star. The
TESS photometry eventually revealed the presence of a possible
transiting terrestrial planet with a short orbital period of
$\sim$2~days, and so the planet candidate was designated as a TESS
Object of Interest (TOI) \citep{guerrero2021a}. These additional data
sources and tentative planet detections enable an opportunity to
revisit the HD~104067 system, and determine the dynamical consequences
of the revised planetary system architecture.

Here, we present a detailed analysis of the HD~104067, including RV
and photometric data, that refines the properties of the known planet,
reveals the presence of an additional RV planet, and provides updated
properties for the transiting planet candidate. We further conduct a
dynamical analysis of the three planet model, investigating the
orbital stability and possible tidal consequences for the inner
planet. Section~\ref{obs} describes the observational data, including
their extraction and preparation for analysis. Section~\ref{analysis}
provides the results of our analysis of the data, and the full
architecture model for the system. In Section~\ref{dynamics}, we
present the results of a detailed dynamical analysis, demonstrating
that it is long-term stable and that planet-planet interactions may
result in significant tidal effects for the inner
planet. Section~\ref{discussion} discusses the implications of the
system architecture and follow-up observations, and we provide
concluding remarks in Section~\ref{conclusions}.

%%%%%%%%%%%%%%%%%%%%%%%%%%%%%%%%%%%%%%%%%%%%%%%%%%%%%%%%%%%%%%%%%%%%

\section{Observations}
\label{obs}

Here we briefly describe the two main data sources used in this
analysis: the RVs from HARPS and HIRES and the photometric data from
TESS.

%%%%%%%%%%%%%%%%%%%%%%%%%%%%%%%%%%%%%%%%%%%%%%%%%%%%%%%%%%%%%%%%%%%%

\subsection{Radial Velocities}
\label{rvobs}

The original discovery of planet b orbiting HD~104067 by
\citet{segransan2011} employed 88 RV observations obtained using the
HARPS spectrograph. Since then, HARPS ceased monitoring of the target
and the HARPS temporal coverage of this target remains around 2271
days from February 2004 to April 2010. In 2020, a re-reduction of all
HARPS data carried out by \citet{trifonov2020a} corrected systematics,
such as nightly zero-point RVs and intra-night drifts, that slightly
improved the overall HARPS precision. These revised HARPS reductions
were utilized for our subsequent analysis, and the HD~104067 data
yield a mean RV uncertainty of $\sim$0.89~m/s. In addition, the Keck
RV program has been carrying out observations of HD~104067 for over
two decades using the HIRES spectrograph, and were used in an
independent analysis of the system by \citet{rosenthal2021}. In total,
72 HIRES RVs were collected, spanning from January 1997 to May 2023
($\sim$9600~days) with an average RV uncertainty of $\sim$1.28~m/s.

%%%%%%%%%%%%%%%%%%%%%%%%%%%%%%%%%%%%%%%%%%%%%%%%%%%%%%%%%%%%%%%%%%%%

\subsection{TESS Photometry}
\label{phot}

HD~104067 has been observed during TESS sectors 10, 36, and 63. To
search for signs of stellar activity, we utilized the 2-minute cadence
Simple Aperture Photometry (SAP) light curves that have been processed
by the Science Processing Operations Center
\citep[SPOC;][]{jenkins2016} and are available on the Mikulski Archive
for Space Telescopes (MAST):
\dataset[10.17909/t9-nmc8-f686]{https://doi:10.17909/t9-nmc8-f686}. Data
points flagged as poor quality were removed, in addition to 5$\sigma$
outliers. A Lomb-Scargle periodogram was used to estimate a stellar
variability period of $18.3\pm4.9$~days from each single-sector light
curve. We further discuss a possible interpretation of the observed
variability in Section~\ref{stellar}. To constrain the planet
candidate's orbital and physical parameters, we used the Pre--search
Data Conditioning (PDC) lightcurves, since the flux measurements are
adjusted for dilution from nearby stars. We describe the preparation
of the lightcurve and fitting process in Section~\ref{tessplanet}.

Similar to the Kepler spacecraft, TESS is subject to known observing
and instrumental systematics during observations
\citep{jenkins2016}. These include fluctuations in pauses in data
collection after each orbit, changes in flux due to temperature
changes, and periodic pointing corrections. While the pipeline that
produces the PDCSAP light curves corrects these known systematics in
the majority of TESS targets, some astrophysical changes in flux can
also be removed. In particular, changes in flux that are on the order
of the length of an individual sector tend to be flattened in the
PDCSAP light curves. We find this to be the case for HD~104067, and
thus have selected to alternatively analyze the SAP light curves. Some
known systematics persist, such as changes in flux due to temperature
fluctuations near gaps in the observations, but the long-period
variations are sufficiently large in amplitude and persist over a
multi-year baseline such that we consider these variations to be
physically associated with the star.

%%%%%%%%%%%%%%%%%%%%%%%%%%%%%%%%%%%%%%%%%%%%%%%%%%%%%%%%%%%%%%%%%%%%

\section{Observational Data Analysis}
\label{analysis}

Here we provide the results from our analysis of the RV and
photometric data, including extracted parameters for the star and
planetary candidates.

%%%%%%%%%%%%%%%%%%%%%%%%%%%%%%%%%%%%%%%%%%%%%%%%%%%%%%%%%%%%%%%%%%%%

\subsection{Stellar Properties}
\label{stellar}

\begin{deluxetable*}{lccr}
\tablecaption{\label{tab:star} HD~104067 derived stellar parameters.}
\tablehead{\colhead{~~~Parameter} & \colhead{Units} & \colhead{Values} & \colhead{Source}}
\startdata
~~~~$V$\dotfill & V-band magnitude\dotfill & $7.93$ & \\
~~~~$M_\star$\dotfill & Mass ($M_\odot$)\dotfill & $0.82\pm0.03$ & This work, Specmatch\\
~~~~$R_\star$\dotfill & Radius ($R_\odot$)\dotfill & $0.78\pm0.01$ & This work, Specmatch\\
~~~~$\log{g}$\dotfill & Surface gravity (cgs)\dotfill & $4.56\pm0.10$ & This work, Specmatch\\
~~~~$T_{\rm eff}$\dotfill & Effective Temperature (K)\dotfill & $4952\pm100$ & This work, Specmatch\\
~~~~$[{\rm Fe/H}]$\dotfill & Metallicity (dex)\dotfill & $0.11\pm0.06$ & This work, Specmatch\\
~~~~$v \sin i$\dotfill & Projected rotational velocity (km/s)\dotfill & $2.32\pm1.0$ & This work, Specmatch\\
~~~~$P_{\rm rot}$\dotfill & Stellar rotation period (days)\dotfill & $18.3\pm4.9$ & This work, TESS\\
~~~~$\varpi$\dotfill & Parallax (mas)\dotfill & $49.1470\pm0.0235$ & Gaia DR3\\
~~~~$d$\dotfill & Distance (pc)\dotfill & $20.35\pm0.01$ & Gaia DR3\\
\enddata
\end{deluxetable*}

Given the brightness of HD~104067, the star is also known by numerous
aliases, such as GJ~1153, HIP~58451, TIC~428673146, and Gaia DR3
3494677900774838144. \citet{segransan2011} refer to HD~104067 as a
moderately active K2V star with a spectroscopically determined
rotation period of 34.7~days. We derived stellar properties for the
star by applying the {\sc SpecMatch} \citep{petigura2015} and {\sc
  Isoclassify} \citep{huber2017d} software packages to the obtained
template Keck-HIRES spectrum. These parameters are provided in
Table~\ref{tab:star}. In addition, we include in Table~\ref{tab:star}
the parallax and distances extracted from the third data release of
the Gaia mission \citep{brown2021}. In addition to the discussion of
stellar properties in the context of the known exoplanet detection,
provided by \citet{segransan2011} and \citet{rosenthal2021}, the star
has also been the subject of other surveys. For example,
\citet{suarezmascareno2015} spectroscopically determined a rotation
period of $29.8\pm3.1$~days, roughly consistent with that found by
\citet{segransan2011}.

\begin{figure*}
    \centering
    \includegraphics[width=1.0\linewidth]{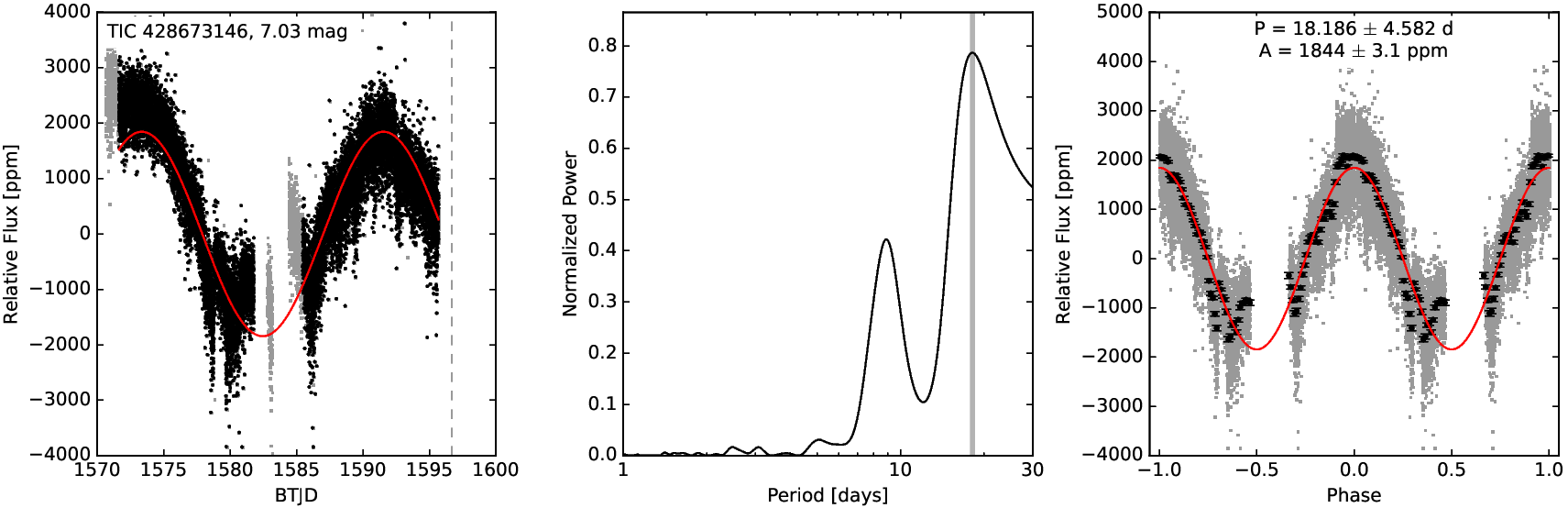}
    \includegraphics[width=1.0\linewidth]{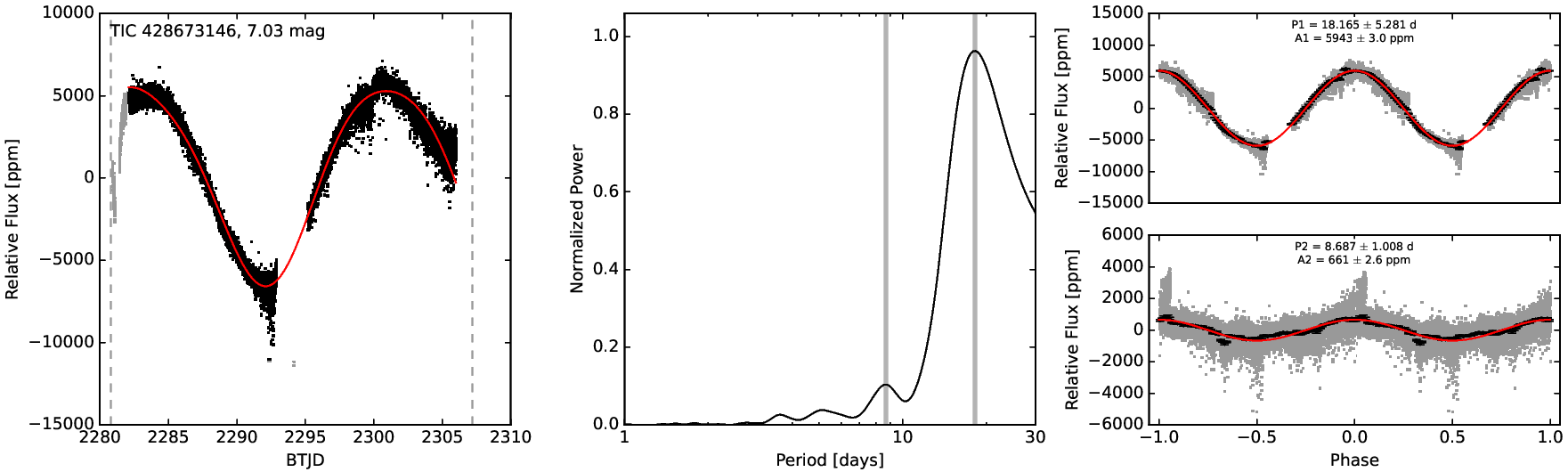}
    \includegraphics[width=1.0\linewidth]{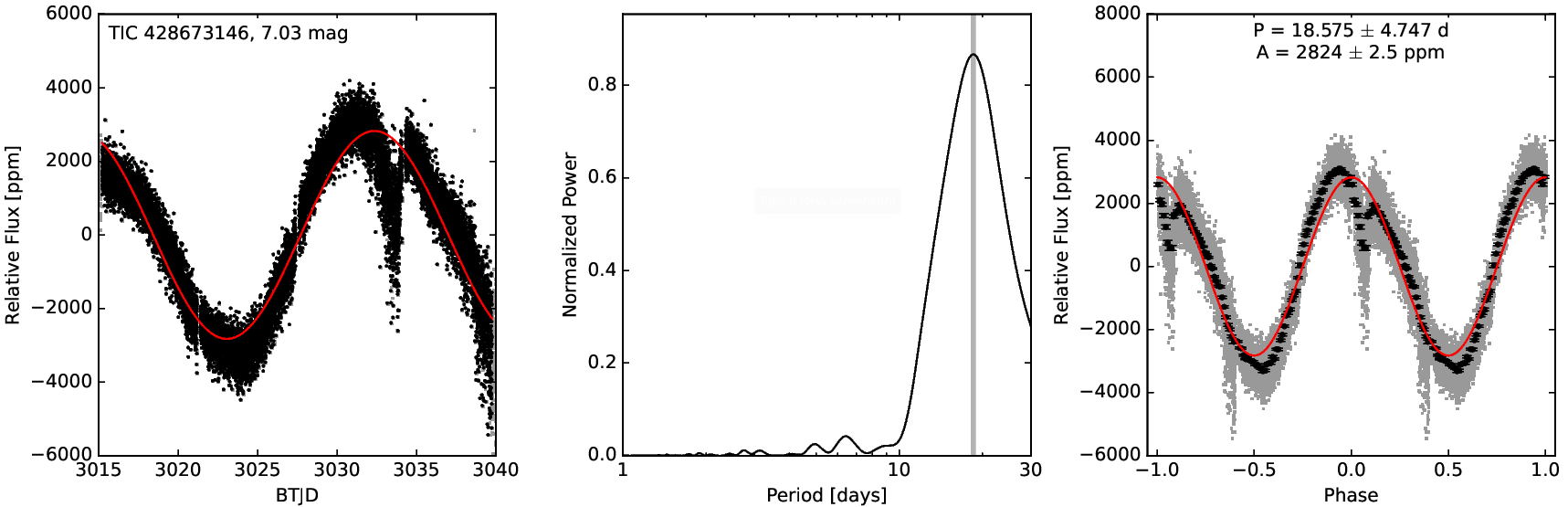}
  \caption{TESS 2-minute cadence photometry (left column),
    Lomb-Scargle periodogram (center column), and light curve
    phase-folded on the measured stellar rotation period (right
    column) for sectors 10 (top row), 36 (center row), and 63 (bottom
    row). The gray points in the left column panels show data points
    that were removed based on poor quality flags or 5$\sigma$
    outliers. The red curve represents the sinusoidal function that
    best-fits the light curve and represents the measured stellar
    rotation period. The black points in the right column represent
    the binned photometry.}
  \label{fig:variability}
\end{figure*}

As mentioned in Section~\ref{phot}, we used a Lomb-Scargle periodogram
to examine stellar variability that may be associated with the
rotation period of the star from each individual sector of TESS
photometry. Shown in Figure~\ref{fig:variability} are the photometry,
periodogram analysis, and folded lightcurves for TESS sectors 10 (top
row), 36 (middle row), and 63 (bottom row). These analyses reveal a
consistent periodic variability signature of $18.3\pm4.9$~days for the
three sectors, despite the observations each being separated by
$\sim$1~year. The second period identified in sector 36 is
approximately half of the primary variability period, and thus is
likely an alias. If the 18.3~day signal is indeed the signature of the
stellar rotation period, then it is quite different from that derived
spectroscopically by \citet{segransan2011} and
\citet{suarezmascareno2015}. A possible cause of this discrepancy in
rotation period measurements is differential rotation, similar to that
observed for the known exoplanet host HD~219134
\citep{johnson2016b}. Furthermore, there is growing evidence that
spectroscopic and photometric rotation period measurements do not
always agree, possibly the result of activity differences in the
chromosphere and photosphere
\citep{lafarga2021,schofer2022,isaacson2024}. However, there are
several other pieces of information to note here. First, the $R_\star$
and $v \sin i$ values shown in Table~\ref{tab:star} result in a
calculated rotation period of $\sim$17~days, which is more consistent
with the rotation period derived from TESS photometry than those
previously determined spectroscopically. \citet{segransan2011}
measured a $v \sin i = 1.61$~km/s; slower than the 2.32~km/s value
shown in Table~\ref{tab:star}, and enough to account for their larger
rotation period of $\sim$30~days. Second, the TESS photometry is
substantially separated in time from the HARPS RV data, such that the
star may be located at a significantly different place within its
magnetic activity cycle, as has been photometrically observed for
numerous other stars \citep{henry2000d,dragomir2012a}. Third, it is
worth acknowledging that aliases of the true rotation period can be
difficult to disentangle, depending on the sampling rate of the
observations, and so some of the reported rotation periods may be
examples of such aliases.

%%%%%%%%%%%%%%%%%%%%%%%%%%%%%%%%%%%%%%%%%%%%%%%%%%%%%%%%%%%%%%%%%%%%

\subsection{Non-Transiting Planets}
\label{rvplanets}

%%%%%%%%%%%%%%%%%%%%%%%%%%%%%%%%%%%%%%%%%%%%%%%%%%%%%%%%%%%%%%%%%%%%

\subsubsection{Radial Velocity Solutions}
\label{rvsol}

We utilized both HARPS and HIRES data in our RV analysis. In our
modeling process, we divided the HIRES dataset into two separate
datasets (HIRES1 and HIRES2) due to a major upgrade carried out on
HIRES in August 2004 \citep{butler2017} that could introduce a
velocity offset between the two datasets. We first used an iterative
planet search tool, \texttt{RVSearch}, to look for potential planetary
signals within the combined data. We refer readers to
\citet{rosenthal2021} for details regarding the working process of
\texttt{RVSearch}. We set the algorithm to search within a period
space between 2 days and five times the combined observing baseline of
HARPS and HIRES, which is around 48,000 days. The search process
includes instrument offsets as well as linear and quadratic trends,
and we only considered periodogram signals that have a false alarm
probability (FAP) less than 0.1\%, as described by \citet{howard2016}
and \citet{fulton2018a}. \texttt{RVSearch} returned two significant
signals from the search: one belonging to the known b planet at
$\sim$58~days with an FAP that is consistent with zero, and the other
yielding a new period signature of $\sim$14~days with an FAP of
$3.29\times10^{-7}$. It is worth noting that the 14~day signal is
primarily detectable within the HARPS data, which is unsurprising
given the high cadence nature of the HARPS observations as well as
their slightly higher RV precision compared to the HIRES data.

\begin{figure*}
  \begin{center}
     \includegraphics[width=14.0cm]{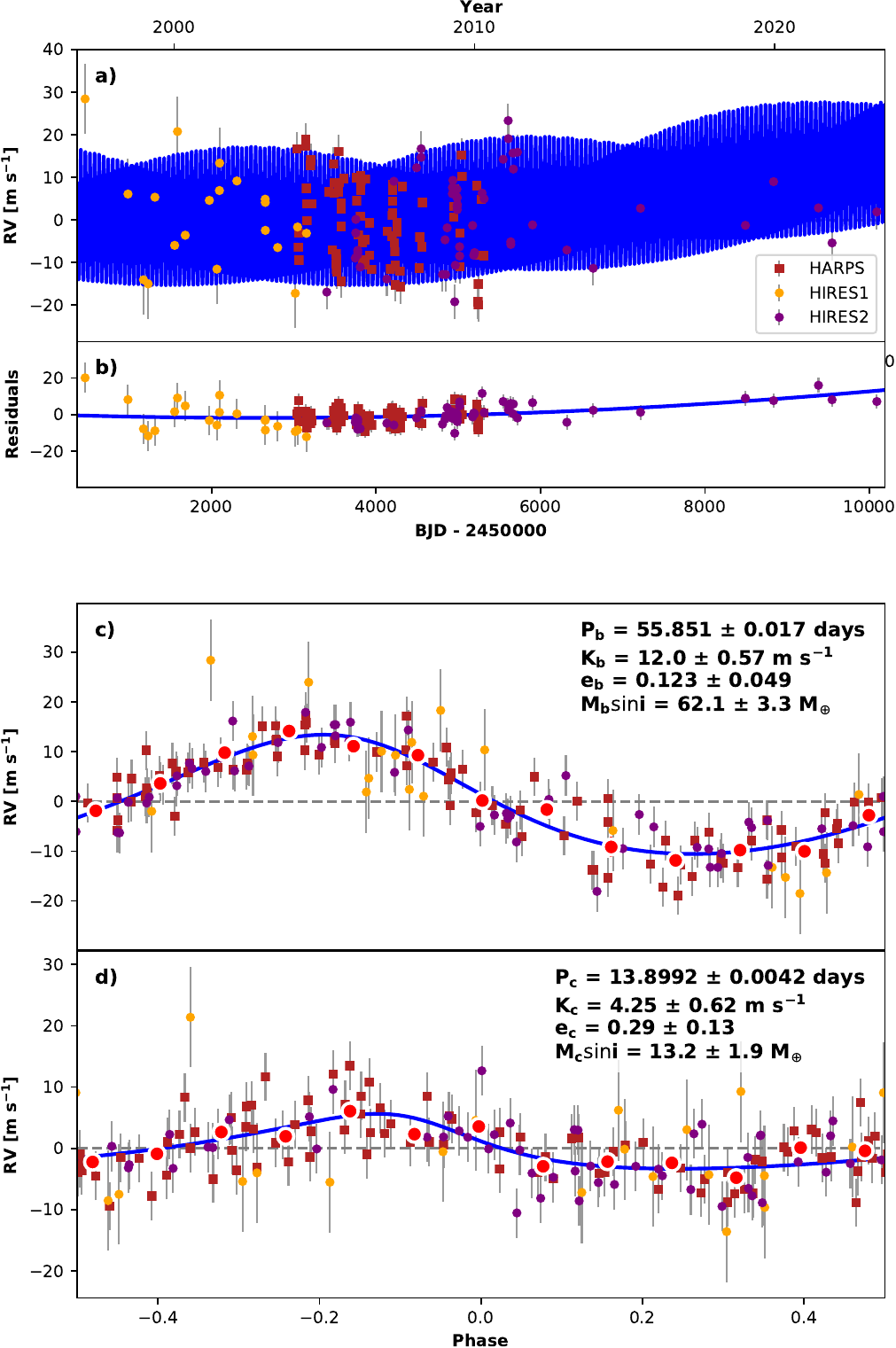}
  \end{center}
  \caption{RV data for HD~104067, including those acquired from HARPS
    and HIRES on either side of the 2004 HIRES upgrade (HIRES1 and
    HIRES2), where the blue lines indicate the best fit to the
    data. The top panels show the full observational baseline and fit
    to the residuals. The bottom panels show the data fits for the
    individual planets, b and c.}
  \label{fig:rvs}
\end{figure*}

The \texttt{RVSearch} results were then used as input for the RV
modeling toolkit \texttt{RadVel} \citep{fulton2018a} to sample
posteriors of Keplerian orbital parameters of the returned signals, as
well as to estimate the associated uncertainties via a Markov Chain
Monte Carlo (MCMC) analysis. All parameters, including instrument
offset and trends, were allowed to vary as free parameters. Chain
convergences were evaluated under four criteria: Gelman-Rubin
statistic ($<0.01$), minimum autocorrelation time factor ($>40$),
maximum relative change in autocorrelation time ($<0.03$), and minimum
independent draws ($>1000$). The MCMC chain rapidly converged, and we
present the full results of our two planet model in
Table~\ref{tab:rvplanets}. The planetary masses and semi-major axes
were derived using the stellar mass shown in Table~\ref{tab:star}. The
model includes both a linear and a quadratic trend, with significances
of 3$\sigma$ and 2$\sigma$, respectively. The RV data are shown in
Figure~\ref{fig:rvs}, where the blue lines indicate the best fit to
the data. The top panels show the full observational baseline and the
residuals from the fit to the data. The bottom panels show the
individual fits to the two planetary signatures detected in the RV
data: the known 55.8~day period planet (planet b), and the newly
detected 13.9~day period planet (planet c).

\begin{deluxetable*}{lrrr}
\tablecaption{\label{tab:rvplanets} Radial velocity derived planetary parameters.}
\tablehead{
  \colhead{Parameter} & 
  \colhead{Credible Interval} & 
  \colhead{Maximum Likelihood} & 
  \colhead{Units}
}
\startdata
\sidehead{\bf{Orbital Parameters}}
  $P_{b}$ & $55.851\pm 0.017$ & $55.851$ & days \\
  $T\rm{conj}_{b}$ & $2454167.3^{+0.94}_{-0.87}$ & $2454167.0$ & BJD \\
  $T\rm{peri}_{b}$ & $2454159.5^{+3.5}_{-3.4}$ & $2454159.7$ & BJD \\
  $e_{b}$ & $0.123^{+0.048}_{-0.051}$ & $0.136$ &  \\
  $\omega_{b}$ & $0.5^{+0.4}_{-0.41}$ & $0.5$ & radians \\
  $K_{b}$ & $12.0\pm 0.57$ & $11.98$ & m s$^{-1}$ \\
  $P_{c}$ & $13.8992^{+0.0047}_{-0.0037}$ & $13.8985$ & days \\
  $T\rm{conj}_{c}$ & $2455192.9^{+0.59}_{-0.56}$ & $2455192.64$ & BJD \\
  $T\rm{peri}_{c}$ & $2455191.6^{+1.2}_{-1.1}$ & $2455191.6$ & BJD \\
  $e_{c}$ & $0.29^{+0.12}_{-0.13}$ & $0.32$ &  \\
  $\omega_{c}$ & $0.64^{+0.57}_{-0.56}$ & $0.66$ & radians \\
  $K_{c}$ & $4.25^{+0.62}_{-0.63}$ & $4.49$ & m s$^{-1}$ \\
\hline
\sidehead{\bf{Other Parameters}}
  $\gamma_{\rm HIRES2}$ & $-1.81^{+0.72}_{-0.74}$ & $-1.59$ & m s$-1$ \\
  $\gamma_{\rm HIRES1}$ & $0.0^{+2.6}_{-2.8}$ & $0.4$ & m s$-1$ \\
  $\gamma_{\rm HARPS}$ & $1.02^{+0.84}_{-0.82}$ & $1.14$ & m s$-1$ \\
  $\dot{\gamma}$ & $0.0013^{+0.00042}_{-0.00043}$ & $0.00142$ & m s$^{-1}$ d$^{-1}$ \\
  $\ddot{\gamma}$ & $2.7e-07\pm 1.3e-07$ & $2.7e-07$ & m s$^{-1}$ d$^{-2}$ \\
  $\sigma_{\rm HIRES2}$ & $4.34^{+0.6}_{-0.52}$ & $3.85$ & $\rm m\ s^{-1}$ \\
  $\sigma_{\rm HIRES1}$ & $8.2^{+1.1}_{-1.2}$ & $8.1$ & $\rm m\ s^{-1}$ \\
  $\sigma_{\rm HARPS}$ & $4.05^{+0.37}_{-0.33}$ & $3.83$ & $\rm m\ s^{-1}$ \\
\hline
\sidehead{\bf{Derived Posteriors}}
  $M_b\sin i$ & $62.1^{+3.3}_{-3.2}$ & $61.5$ & M$_{\oplus}$ \\
  $a_b$ & $0.2674^{+0.0032}_{-0.0033}$ & $0.2671$ &  AU \\
  $M_c\sin i$ & $13.2\pm 1.9$ & $13.2$ & M$_{\oplus}$ \\
  $a_c$ & $0.1058\pm 0.0013$ & $0.1057$ &  AU \\
\enddata
\tablenotetext{}{80000 links saved}
\tablenotetext{}{
  Reference epoch for $\gamma$,$\dot{\gamma}$,$\ddot{\gamma}$: 2455275.946932 
}
\end{deluxetable*}

%%%%%%%%%%%%%%%%%%%%%%%%%%%%%%%%%%%%%%%%%%%%%%%%%%%%%%%%%%%%%%%%%%%%

\subsubsection{Planet Search Completeness}

\begin{figure*}
  \begin{center}
     \includegraphics[width=14.0cm]{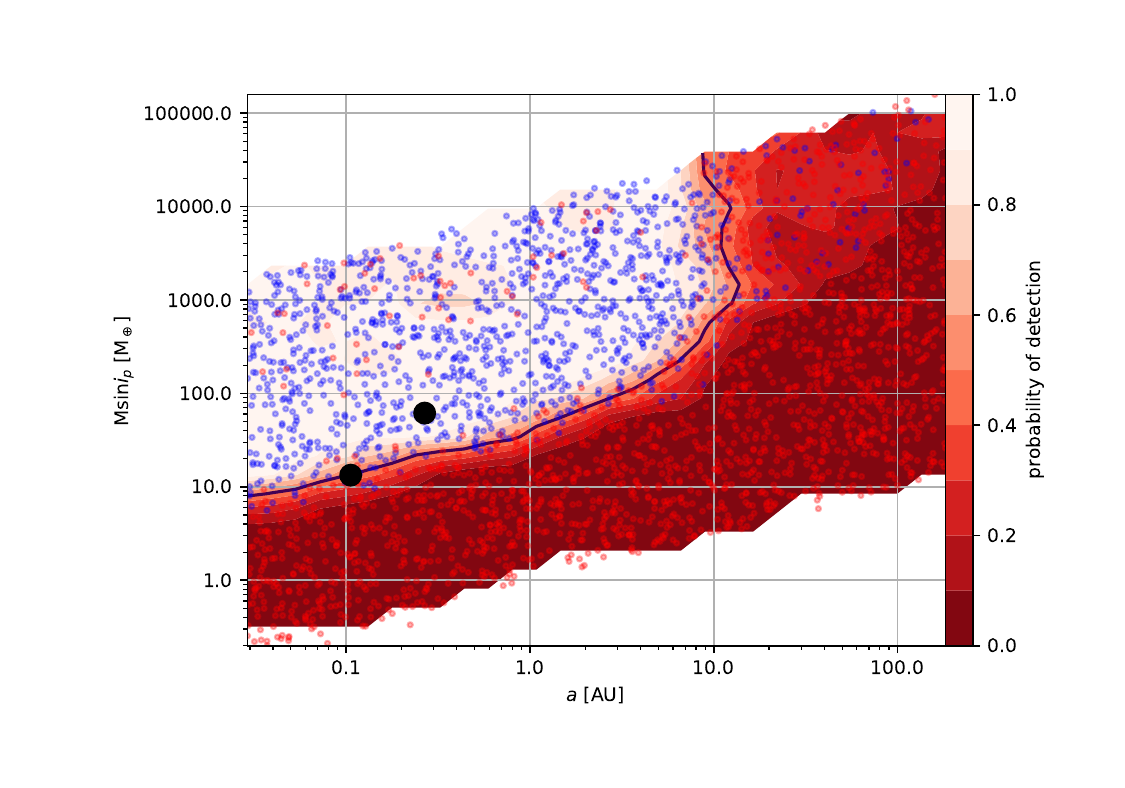}
  \end{center}
  \caption{Results of the injection-recovery test to determine the
    sensitivity of the RV data to planetary signatures as a function
    of planetary mass ($M_p \sin i$) and semi-major axis ($a$). The
    blue dots represent injected planetary signatures that were
    successfully recovered and the red dots represent those planets
    that were not recovered. The color scale shown on the right
    vertical axis corresponds to the probability contours of detecting
    a planet of a given mass and semi-major axis. The two detected
    planets are shown as large black dots.}
  \label{fig:injection}
\end{figure*}

We carried out an injection and recovery test on the combined HARPS
and HIRES RV data to examine the sensitivity of RV data to potential
additional companions in the system. In total, 3000 synthetic planets
were injected into the data, with orbital eccentricities drawn from a
beta distribution using parameters from \citet{kipping2013b}, and
orbital periods and minimum masses drawn from log-uniform
distributions. We then used the results from the injection and
recovery test to compute a search completeness contour, shown in
Figure~\ref{fig:injection} as a function of mass ($M_p \sin i$) and
semi-major axis ($a$). The two recovered signals described in
Section~\ref{rvsol} are marked as solid black dots with the
$\sim$14-day signal lying on the 50\% completeness curve, indicated by
the black line. The individual blue and red dots indicate whether the
recovery was successful at each injected location. The results suggest
that additional giant planets more massive than the already discovered
planet b are clearly recoverable if present in the inner part of the
system, but were not detected within our data, suggesting that such an
additional planetary presence is extremely unlikely within 1--2~AU of
the host star. However, our data do not rule out their presence beyond
10~AU, which corresponds roughly to the observing baseline of the
combined RV dataset. On the other hand, smaller planets less massive
than 10 Earth masses lie below the 50\% completeness curve for the
majority of the parameter space, hinting the possibility that small
planets may yet be present within the system. These small planets sit
below the sensitivity of our RV data and their discoveries require
further observations and/or more precise measurements. Overall, our
injection-recovery analysis is able to exclude Saturn-mass and
Jupiter-mass planets out to 3~AU and 8~AU, respectively. It is worth
noting that the completeness analysis described here refers to $M_p
\sin i$, and so depends on the inclination of the synthetic planetary
orbits relative to the plane of the sky.

%%%%%%%%%%%%%%%%%%%%%%%%%%%%%%%%%%%%%%%%%%%%%%%%%%%%%%%%%%%%%%%%%%%%

\subsubsection{Stellar Activity}

Periodic stellar activity cycles, such as rotation, long term magnetic
cycles, and their aliases, can mimic RV signatures similar to those
induced by exoplanets and therefore sometimes could be mistakenly
identified as exoplanet candidates
\citep{desort2007,robertson2014a,kane2016a}. As a crucial step in all
exoplanet discovery and follow-up work, it is essential to check
whether activity could be the primary source contributing to the
periodicities identified in the RV time series. Here, we make use of
all available stellar activity indicators for both HARPS and HIRES to
check if any of the activity periods coincide with our identified RV
signals in Section~\ref{rvsol}. For HARPS, we used all available
activity indicators provided within the HARPS RV database
\citep{trifonov2020a}, namely the H$\alpha$ index, chromatic index,
differential line width, Na I D index, Na II D index, cross
correlation function (CCF) bisector inverse slope span, CCF full width
at half maximum, and CCF contrast. In addition, we include Ca II H \&
K measurements from the second version of the HARPS RV database
\citep{perdelwitz2024}. For HIRES, the Ca II H \& K time series serve
as the only activity indicator. All activity time series were passed
through a Generalized Lomb-Scargle periodogram
\citep{zechmeister2009a} to check for periodic signals within the
observing baseline. No peak in the activity periodograms were found to
be located on or near the periodicities of the two recovered RV
signals described in Section~\ref{rvsol}. These results suggest that
the planetary nature of the newly recovered $\sim$14-day signal is
likely genuine.

%%%%%%%%%%%%%%%%%%%%%%%%%%%%%%%%%%%%%%%%%%%%%%%%%%%%%%%%%%%%%%%%%%%%

\subsection{An Inner Transiting Planet Candidate}
\label{tessplanet}

As described in Section~\ref{phot}, the TESS photometry consists of
three sectors: 10, 36, and 63. For the analysis of the transit signal,
designated as TOI-6713.01, we used the \texttt{LightKurve} package
\citep{lightkurve2018} to download the relevant sectors of the PDC and
SAP time series measurements from MAST. We prepared the data by first
using a Savitzky--Golay filter \citep{savitzky1964} to temporarily
flatten the light curve after masking out the transits (with one
transit duration on either side). An outlier mask was created by
performing 5 iterations of 3$\sigma$ clipping on the flattened light
curve. We then also removed segments of the light curve that were
heavily affected by momentum dumps. This, in addition to the outlier
mask, removed $\sim$8\% of the light curve data.

\begin{figure}
  \includegraphics[width=8.5cm]{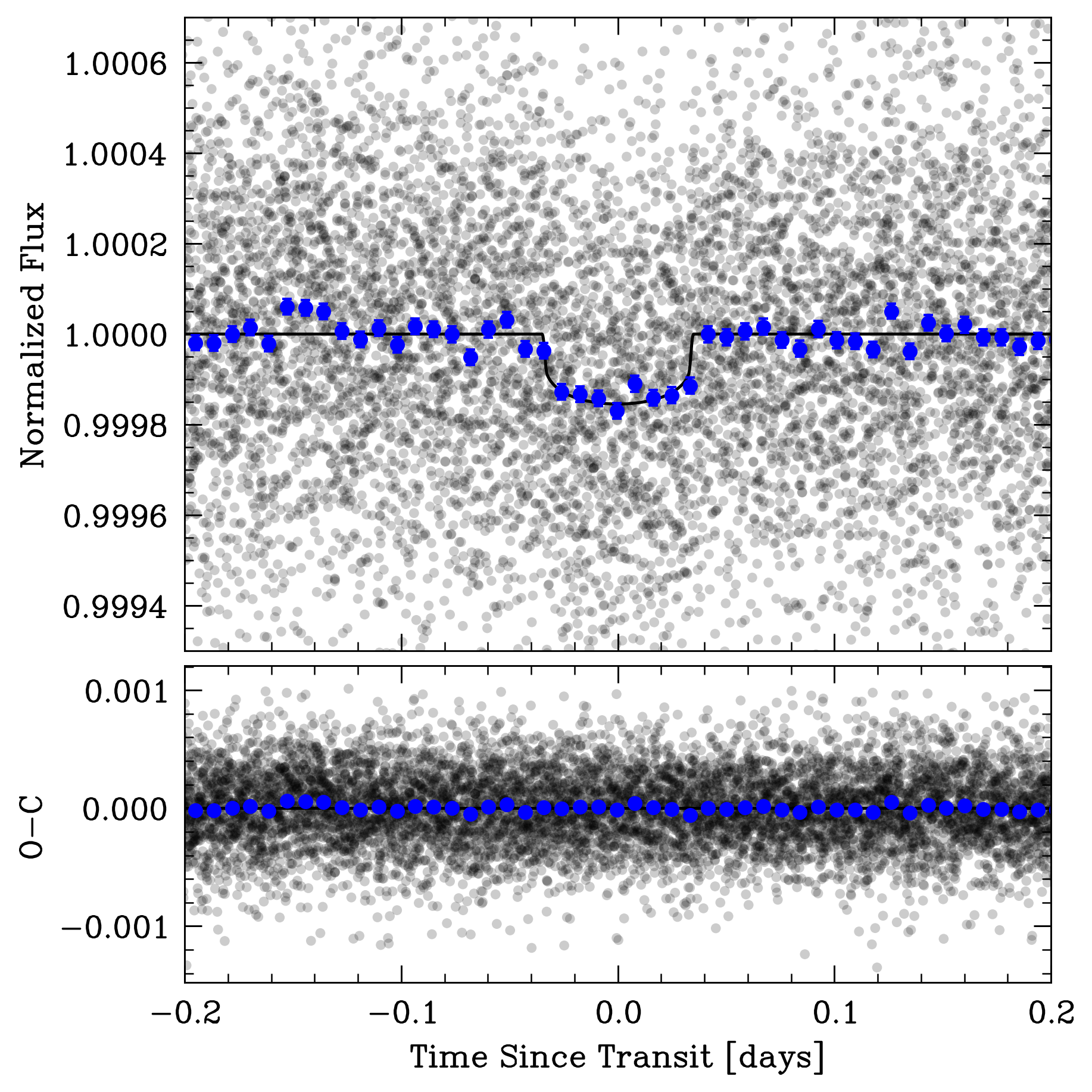}
  \caption{Folded TESS light curve, where gray points are the raw data
    and blue represent the binned data. The median transit model is
    plotted as the black line. Residuals (data minus model) are shown
    in the lower panel.}
  \label{fig:tess_fold}
\end{figure}

\begin{deluxetable*}{lrr}
\tablecaption{\label{tab:tess_params} TOI-6713.01 TESS Transit Parameters}
\tablehead{
  \colhead{Parameter [units]} & 
  \colhead{Credible Interval} & 
  \colhead{Source}
}
\startdata
\sidehead{\bf{Transit Parameters}}
  $P$ [days] & $2.1538197\pm 0.0000041$ & This work, fit \\
  $T\rm{conj}$ [BJD$_{\text{TDB}}-2457000.0$] & $1571.52211\pm0.00081$ & This work, fit \\
  $R_p/R_\star$ [--] & $0.0152\pm0.0014$  & This work, fit \\
  $T_{14}$ [hours] & $1.667\pm0.065$  & This work, fit \\
  $b$ [--] & $0.52\pm 0.29$  & This work, fit \\
  $e$ [--] & $\equiv 0$  & This work \\
  $\omega$ [--] & $\equiv 0$  & This work \\
  $R_p$ [$R_\oplus$] & $1.30\pm0.12$  & This work, derived \\
  $a/R_\star$ [--] & $8.42\pm0.15$ & This work, derived  \\
  $a$ [AU] & $0.03054\pm0.00037$ & This work, derived  \\
  $i$ [deg] & $86.5\pm2.0$ & This work, derived \\
\hline
\sidehead{\bf{Gaussian Process Parameters}}
  $\ln{(Q)}$ [--] & $-6.59\pm 0.53$ & This work, fit \\
  $\ln{(dQ)}$ [--] & $-1.43\pm 0.75$ & This work, fit \\
  $P_{\text{rot}}$ [days] & $8.9\pm 2.6$ & This work, fit \\
  $\ln{(\sigma)}$ [--] & $-8.30\pm 0.12$ & This work, fit \\
  $f$ [--] & $0.093\pm 0.025$ & This work, fit \\
\hline
\enddata
\tablenotetext{}{}
\tablenotetext{}{}
\end{deluxetable*}

We fit the prepared light curve using the \texttt{exoplanet} package
\citep{foremanmackey2021}, which uses a Hamiltonian Monte Carlo (HMC)
routine to explore posterior probability distributions. To model the
transit, we assumed a circular orbit with orbital period ($P$), time
of inferior conjunction ($T_\mathrm{conj}$), scaled planet radius
($R_p/R_\star$), impact parameter ($b$), transit duration ($T_{14}$),
and mean flux offset ($\mu$) as free parameters. Quadratic limb
darkening coefficients were held constant at $u_0=0.45$ and $u_1=0.18$
\citep{claret2017a}. Gaussian priors were set on $P$ and
$T_\mathrm{conj}$ based on values reported by the Exoplanet Follow-up
Observing Program (ExoFOP).

The stellar variability was modeled using the RotationTerm kernel
included in \texttt{celerite2} \citep{foremanmackey2017}, which is a
mixture of two simple harmonic oscillators. For the secondary
oscillation quality factor ($Q$), the difference in quality factors
between the primary and secondary modes ($dQ$), fractional amplitude
of secondary mode relative to primary mode ($f$), and the standard
deviation of the GP ($\sigma$), we sampled the logarithmic value of
the parameter. A Gaussian prior was placed on the rotation period
($P_{\text{rot}}$) informed by the stellar activity analysis in
Section~\ref{stellar}. However we note the 18 day signal was not
well-preserved in the PDC lightcurve.

We used the values obtained from minimizing a negative log-likelihood
function as initial positions for two parallel chains and ran the HMC
for 2000 tuning steps and 4000 sampling steps per chain. In
Table~\ref{tab:tess_params} we provide the median values with their
1$\sigma$ uncertainties for each fitted and derived parameter. The
model fit to binned TESS data is shown in
Figure~\ref{fig:tess_fold}. The transit depth is 231~ppm and the
out-of-transit PDCSAP photometry has a standard deviation of
$\sim$300~ppm. As there are $\sim$1300 measurements acquired during
transit, the transit signal-to-noise for this detection is $S/N \sim
27$, according to the methodology described by \citet{pont2006c}.
Note that we examined both the PDC and SAP photometry from the
individual TESS sectors and found that the transit signal was present
in all cases, although the signal becomes less significant in the SAP
data. More follow-up observations are needed to validate and confirm
the planetary nature of the photometric signal.

%%%%%%%%%%%%%%%%%%%%%%%%%%%%%%%%%%%%%%%%%%%%%%%%%%%%%%%%%%%%%%%%%%%%

\section{Orbital Dynamics}
\label{dynamics}

Here we present results of a dynamical analysis of the system,
including tidal impacts for the inner transiting planet candidate.

%%%%%%%%%%%%%%%%%%%%%%%%%%%%%%%%%%%%%%%%%%%%%%%%%%%%%%%%%%%%%%%%%%%%

\subsection{Long-Term Stability}
\label{stability}

Given the eccentric orbits of the RV-detected giant planets, and the
prospect of a further inner terrestrial planet (see
Section~\ref{analysis}), the architecture of the system presents an
interesting opportunity to explore planetary dynamical interactions
within a relatively compact system. The dynamical simulations
conducted for this work make use of the Mercury Integrator Package
\citep{chambers1999}, using similar methodology to that conducted by
\citet{kane2019c,kane2021a}. We also adopt the hybrid
symplectic/Bulirsch-Stoer integrator with a Jacobi coordinate system,
which generally provides more accurate results for multi-planet
systems \citep{wisdom1991,wisdom2006b}. We used a time step of
0.05~days (1.2~hours) to ensure adequate sampling of the inner
planetary orbit and ran the simulation for a $10^7$ years. Although
neither of the two RV planets are known to transit, we assumed that
the system is roughly coplanar. Adopting the parameters for the inner
planet described in Section~\ref{tessplanet}, we assumed a circular
orbit for that planet. Given the estimated radius for the inner planet
of $1.30\pm0.12$~$R_\oplus$, we adopted a planetary mass of
2~$M_\oplus$, based on predictions from exoplanet mass-radius
relationships \citep{weiss2014,chen2017}.

\begin{figure*}
  \begin{center}
     \includegraphics[width=16.0cm]{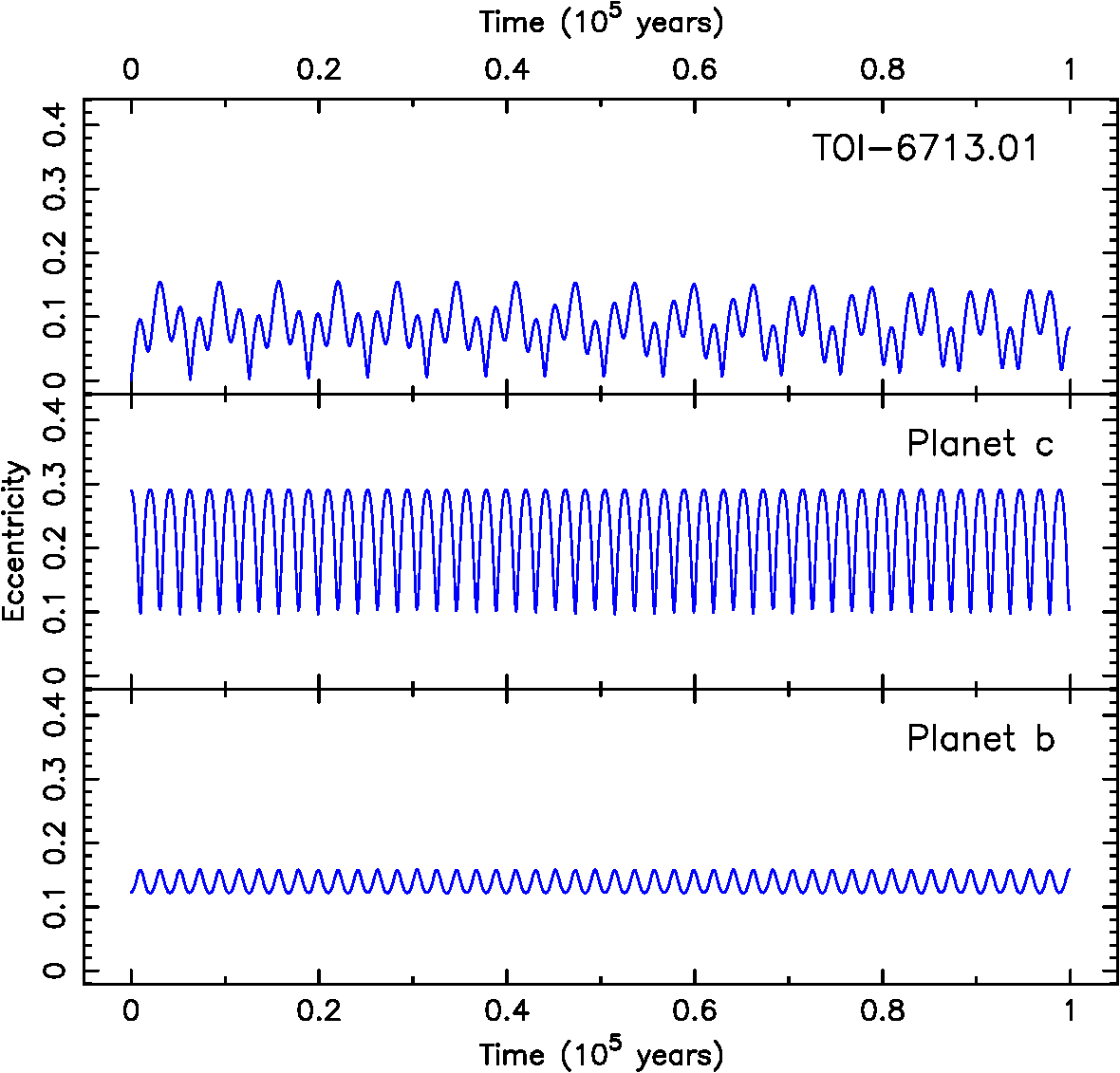}
  \end{center}
  \caption{Eccentricity evolution for the three planet model of the
    HD~104067 system, including the transiting planet candidate,
    TOI-6713.01 (top panel), the newly detected planet c (middle
    panel), and the previously known planet b (bottom panel). Data are
    shown for the first $10^5$ years of the full $10^7$ year
    integration.}
  \label{fig:ecc}
\end{figure*}

Our simulations found that the system is able to retain long-term
stability for the full $10^7$ year integration. The results of our
simulations are represented in Figure~\ref{fig:ecc}, showing the
eccentricity evolution of the transiting planet candidate, TOI-6713.01
(top panel), the newly detected planet c (middle panel), and the
previously known planet b (bottom panel). In order for the structure
of the eccentricity evolution to be easily visible, we have restricted
the data shown in the plot to the first $10^5$ years of the full
$10^7$ year simulation. Note also that we use identical eccentricity
scales for each panel to emphasize the relative amplitudes of the
eccentricity variations. Figure~\ref{fig:ecc} shows that the two giant
planets experience a regular exchange of angular momentum that results
in eccentricity ranges of 0.12--0.16 and 0.10--0.29 for planets b and
c, respectively. The eccentricity range for TOI-6713.01 is 0.00--0.16,
with a clear beating pattern, indicating that the presence of the
inner planet causes a slight difference in frequency between the
angular moment exchange between planets b and c. Since TOI-6713.01
lies so close to the host star, even a relatively small eccentricity
excitation may have significant consequences for tidal forces
experienced by the planet.

%%%%%%%%%%%%%%%%%%%%%%%%%%%%%%%%%%%%%%%%%%%%%%%%%%%%%%%%%%%%%%%%%%%%

\subsection{Tidal Effects}
\label{tides}

Based on the eccentricity evolution resulting from the dynamical
simulations from Section~\ref{stability}, we calculated the tidal
heating of the inner planet candidate and the consequences for the
planetary surface temperature. The power contributed to a rigid body
via tidal heating is given by
\begin{equation}
  P_\mathrm{tides} = \dot{E} = - \mathrm{Im}(k_2) \frac{21}{2}
  \frac{G^{3/2} M_\star^{5/2} R_p^5 e^2}{a^{15/2}}
  \label{equ:tide1}
\end{equation}
where $G$ is the gravitational constant, $M_\star$ is the mass of the
host star, $R_p$ is the radius of the planet, $e$ is the orbital
eccentricity, and $k_2$ is the second-order Love number
\citep{squyres1983b,meyer2007}. For our calculations, we adopt a Love
number of $k_2 = 0.35$, based on the terrestrial bodies
\citep{zhang1992c}. Equation~\ref{equ:tide1} clearly demonstrates the
extreme sensitivity of tidal heating and power output to the
semi-major axis of the body, since it scales with $a^{-15/2}$. For
example, the tidal heating of Io results in a total tidal dissipation
of $10^{14}$~W \citep{bierson2021a}.

The proximity of TOI-6713.01 to the host star will also result in
significant incident flux of stellar radiation that contributes to the
overall energy budget of the planet. The power absorbed by the planet
due to the star is given by
\begin{equation}
  P_\mathrm{star} = \frac{L_\star}{4 \pi a^2} (1-A) \pi R_p^2
  \label{equ:tide2}
\end{equation}
where $L_\star$ is the stellar luminosity and $A$ is the Bond albedo
of the planet. In order to set an upper limit on the stellar
contribution to the planetary heating, we adopt a Bond albedo of $A =
0.0$. For epochs of the simulated dynamical evolution where the planet
has a non-zero eccentricity, we consider only the average flux
received and so evaluate Equation~\ref{equ:tide2} at the semi-major
axis.

To calculate the total power emitted by the planet, we consider the
planet using a blackbody approximation such that the power radiated by
the planet may be expressed as
\begin{equation}
  P_\mathrm{rad} = 4 \pi R_p^2 \epsilon \sigma T_p^4
  \label{equ:tide3}
\end{equation}
where $\epsilon$ is the emissivity of the planet, $\sigma$ is the
Stefan-Boltzmann constant, and $T_p$ is the blackbody equilibrium
temperature of the planet. We assume an emissivity of $\epsilon = 1.0$
which treats the planet as a perfect emitter. As a blackbody, the
power radiated equates to the total power absorbed
($P_\mathrm{star}+P_\mathrm{tides}$). The equilibrium temperature of
the planet is therefore provided by
\begin{equation}
  T_p = \left( \frac{P_\mathrm{star}+P_\mathrm{tides}}{4 \pi R_p^2
    \epsilon \sigma} \right)^{\frac{1}{4}}
  \label{equ:tide4}
\end{equation}
where the power calculations of Equation~\ref{equ:tide1} and
Equation~\ref{equ:tide2} have been incorporated.

\begin{figure*}
  \begin{center}
     \includegraphics[width=16.0cm]{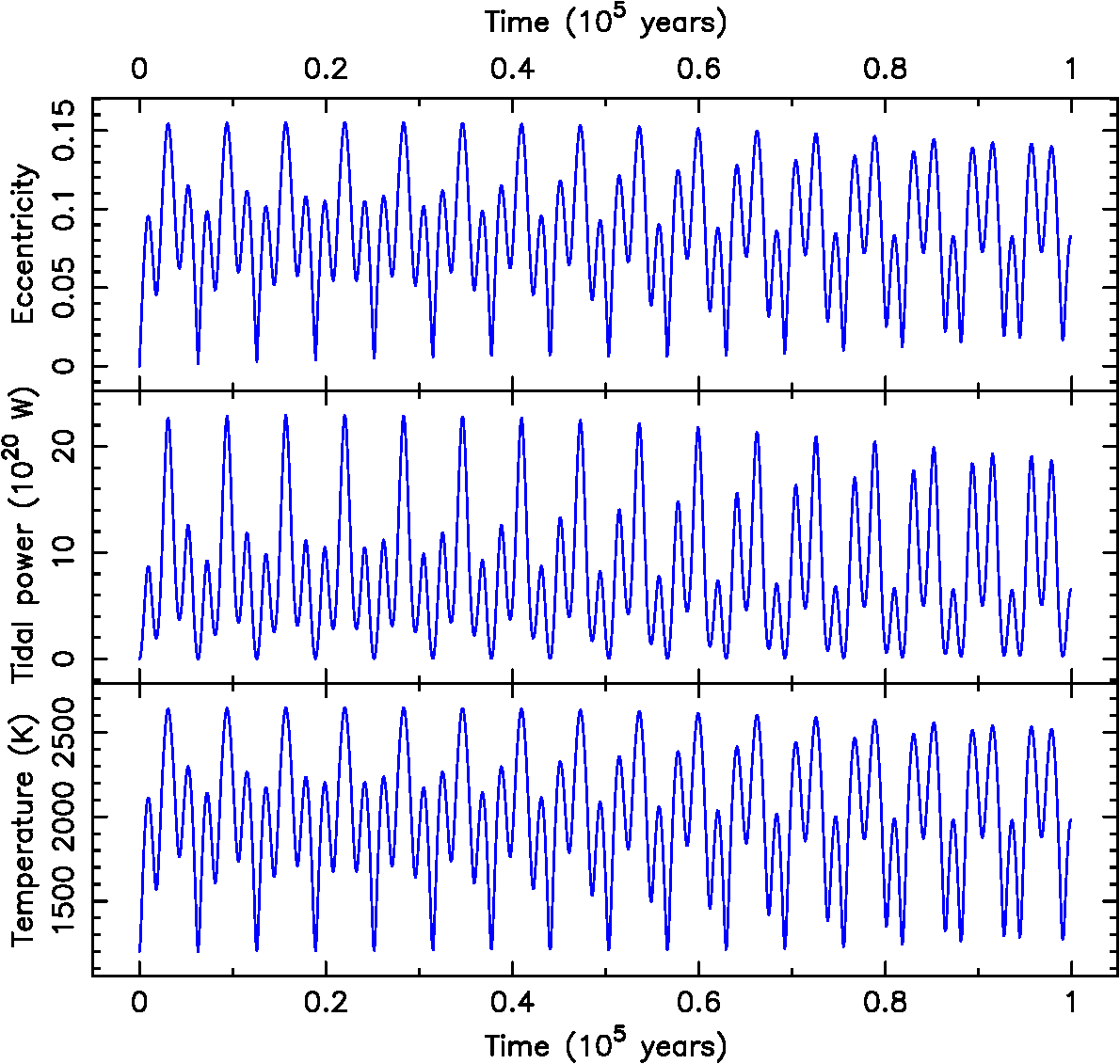}
  \end{center}
  \caption{The eccentricity (top panel), tidal power (middle panel),
    and equilbrium temperature (bottom panel) evolution of the
    transiting planet candidate, TOI-6713.01. As for
    Figure~\ref{fig:ecc}, these results are shown for the first $10^5$
    years of the dynamical simulation.}
  \label{fig:tidal}
\end{figure*}

Figure~\ref{fig:tidal} shows the results of our tidal effects
calculations for TOI-6713.01 for the first $10^5$ years of the
eccentricity evolution (see Section~\ref{stability}). The top panel
shows an expanded view of the eccentricity evolution of the planet
candidate. The middle panel shows the tidal energy dissipation rate
(power) emitted by the planet (Equation~\ref{equ:tide1}). The bottom
panel shows the equilibrium temperature of the planet, combining the
effects of tidal energy and incident stellar flux
(Equation~\ref{equ:tide4}). Clearly, the proximity of the planet to
the host star results in a highly sensitive response of the planet's
tidal energy to the eccentricity variations. The mean tidal power
produced by the planet is $8.3 \times 10^{20}$~W, or almost 7 orders
of magnitude larger than Io. Amazingly, using the above mentioned
assumptions regarding Love number, Bond albedo, and emissivity, these
calculations lead to a very high equilibrium temperature for the
planet, where the temperature varies in the range 1202--2646~K. For an
emissivity of $\epsilon = 0.5$, the equilibrium temperature raises
even higher, reaching peak values of 3130~K. In either scenario, the
peak value of the temperature evolution is well above that needed for
the surface to be in an entirely molten state \citep{boukare2022}.

%%%%%%%%%%%%%%%%%%%%%%%%%%%%%%%%%%%%%%%%%%%%%%%%%%%%%%%%%%%%%%%%%%%%

\section{Discussion}
\label{discussion}

Like many systems that are subjected to detailed follow-up
observations, the HD~104067 system has revealed itself to be
increasingly complex. The additional RV-detected planet (planet c)
provides a potential source for the eccentricity of the previously
known giant planet. The potential for the existence of an inner
terrestrial planet creates a fascinating architecture that is almost
optimal for maximizing the tidal power emission of the
planet. However, TOI-6713.01 has yet to be confirmed. Though the
transiting planet candidate was detected in a known exoplanet system
and the period of the transits is significantly different from the
estimates of intrinsic stellar variability, there remains the
possibility that the transit signal is caused by a false-alarm
scenario. The relatively large pixel sizes of TESS can result in
substantial blending from background stars, creating diluted signals
via blended eclipsing binary stars that can mimic transit signatures
\citep{brown2003a,pont2006c,bryson2013,fressin2013} or effect the
derived planetary radius \citep{ciardi2015a}. A typical pathway to
remedy the risk of false-alarm signals is RV follow-up that is able to
successfully extract a planetary mass \citep{chontos2022a}, and/or the
use of validation steps that search for blend contamination
\citep{giacalone2021}, including high resolution imaging that may
detect nearby stellar companions within the photometric aperture
\citep{howell2011,matson2019,schlieder2021}. Although substantial
efforts toward the imaging of bright exoplanet host stars has been
carried out \citep{kane2014c,kane2015c,wittrock2016,wittrock2017},
much of such efforts have been directed toward hosts of transiting
planets for validation purposes
\citep{law2014,howell2021b}. Furthermore, assuming a planet mass of
2~$M_\oplus$, TOI-6713.01 would have an expected RV amplitude of
$\sim$1~m/s, comparable to the RV precision of both HARPS and HIRES
(see Section~\ref{rvobs}) and significantly below the completeness
contour shown in Figure~\ref{fig:injection}. The RV completeness
results also rule out minimum masses for the transiting planet
candidate greater than $\sim$10~$M_\oplus$, excluding scenarios such
as a grazing stellar companion. Given the brightness of the host star
and the history of previous observations, it is unlikely that stellar
blends are the cause of the transit signature detected by
TESS. However, further observations are recommended to validate the
planetary nature of the transit signal.

\begin{figure}
  \includegraphics[width=8.5cm]{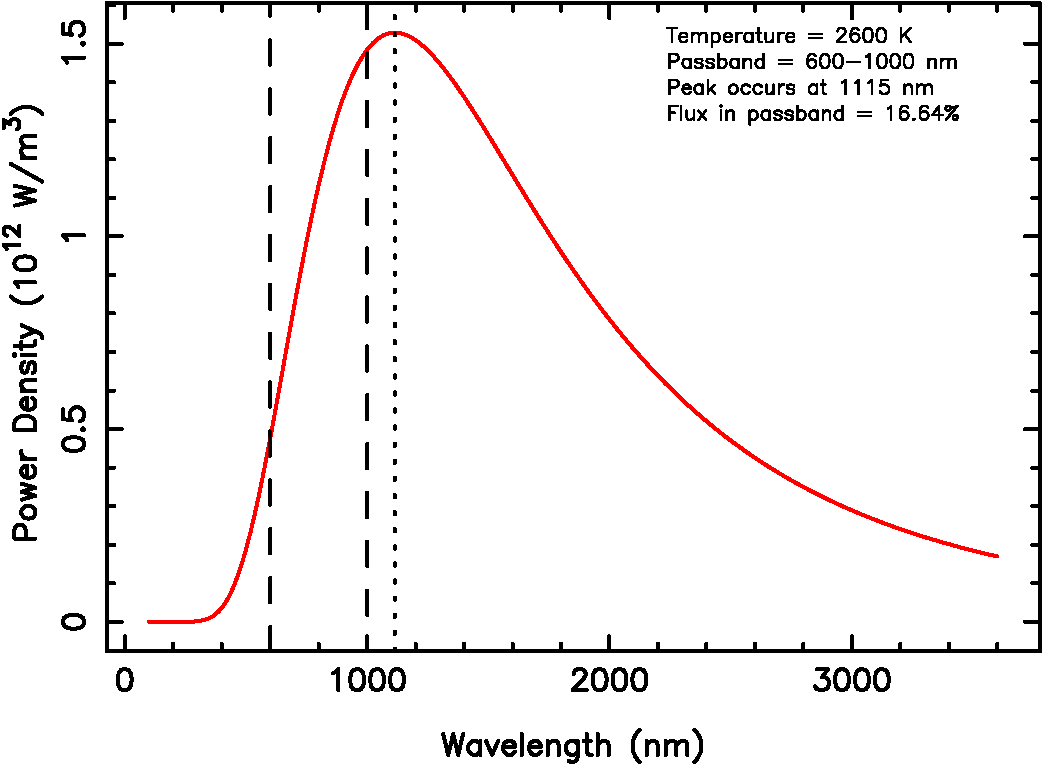}
  \caption{The predicted blackbody flux of TOI-6713.01, assuming a
    calculated temperature of $\sim$2600~K. The passband boundaries of
    TESS are indicated by the vertical dashed lines. Of the integrated
    flux, 16.64\% falls within the TESS passband.}
  \label{fig:blackbody}
\end{figure}

If the transiting planet candidate can be validated, then it may
present an interesting target to determine upper limits on the effects
of tidal heating, described in Section~\ref{tides}. An important
factor in ascertaining the viability of such follow-up observations is
the calculation of predicted flux emission in the context of expected
observational bandpasses. For example, shown in
Figure~\ref{fig:blackbody} is the predicted blackbody spectrum (red
line), assuming the peak equilibrium temperature of 2600~K, as
estimated in Section~\ref{tides}. The vertical lines indicate the
passband boundaries of the TESS instrumentation (600nm--1000nm). Of
the integrated flux from the thermal emission, $\sim$16.64\% of the
total flux falls within the passband of the TESS observations. In
terms of wavelength-dependent flux ratios with the host star, the
expected flux ratios are 83, 92, and 107 ppm at 3.6, 4.5, and 8.0
microns, respectively. The cause of the relatively small amplitude of
these signatures, despite the high equilibrium temperature, is
dominated by the relatively small size of the planet. Note that, as
described in Section~\ref{tides}, the amplitude of these signatures
will vary with time, and often be significantly lower than these peak
values. Moreover, the amplitude of these signatures are below the
measured transit depth (see Section~\ref{tessplanet}) and so will be
challenging to detect within TESS photometry.

%%%%%%%%%%%%%%%%%%%%%%%%%%%%%%%%%%%%%%%%%%%%%%%%%%%%%%%%%%%%%%%%%%%%

\section{Conclusions}
\label{conclusions}

The incredible diversity of observed planetary architectures have also
led to a plethora of dynamical scenarios that deviate substantially
from those observed in the solar system. The opportunities to perform
detailed follow-up studies are often best afforded by systems for
which RV exoplanet detections have been made, since these
preferentially have bright host stars compared with those systems
detected through the transit method. Thus, when transits are detected
in known RV systems, there are often interesting opportunities to
characterize planetary orbits and atmospheres where there has already
been a long history of observations. The HD~104067 system is proving
to be such a case. Our detection of a further giant planet in an
eccentric orbit reveals a compact architecture with a complex
dynamical environment. The potential addition of a short period
terrestrial planet results in an injection of tidal energy into the
planet that has rarely been seen before, and a case study of possible
magma oceans that may have a detectable signature. Our calculations
for the equilibrium temperature of the inner planet candidate lie in
the range 1202–-2646~K, including the effects of extreme stellar flux
and cyclic tidal energy resulting from the eccentricity
evolution. Though the RV and thermal signatures of the inner planet
lie at the threshold of current instruments and facilities, the
continual improvement of measurements for this system may enable
observational tests of the calculations presented here, providing
insight into the formation and evolution of compact planetary systems.

%%%%%%%%%%%%%%%%%%%%%%%%%%%%%%%%%%%%%%%%%%%%%%%%%%%%%%%%%%%%%%%%%%%%

\section*{Acknowledgements}

The authors thank the anonymous referee for their feedback that
improved the manuscript. We gratefully acknowledge the efforts and
dedication of the Keck Observatory staff for support of HIRES and
remote observing. We recognize and acknowledge the cultural role and
reverence that the summit of Maunakea has within the indigenous
Hawaiian community. We are deeply grateful to have the opportunity to
conduct observations from this mountain. This paper includes data
collected by the TESS mission, which are publicly available from the
Mikulski Archive for Space Telescopes (MAST). The results reported
herein benefited from collaborations and/or information exchange
within NASA's Nexus for Exoplanet System Science (NExSS) research
coordination network sponsored by NASA's Science Mission Directorate.

%%%%%%%%%%%%%%%%%%%%%%%%%%%%%%%%%%%%%%%%%%%%%%%%%%%%%%%%%%%%%%%%%%%%

\software{Mercury \citep{chambers1999}, RadVel \citep{fulton2018a},
  LightKurve \citep{lightkurve2018}, exoplanet
  \citep{foremanmackey2021}}

%%%%%%%%%%%%%%%%%%%%%%%%%%%%%%%%%%%%%%%%%%%%%%%%%%%%%%%%%%%%%%%%%%%%

%\bibliographystyle{aasjournal}
%\bibliography{/data/skane/latex/styles/references}

%%%%%%%%%%%%%%%%%%%%%%%%%%%%%%%%%%%%%%%%%%%%%%%%%%%%%%%%%%%%%%%%%%%%

\end{document}